\documentclass[sn-chicago]{sn-jnl}
\usepackage{graphicx}%
\usepackage{multirow}%
\usepackage{amsmath,amssymb,amsfonts}%
\usepackage{amsthm}%
\usepackage{mathrsfs}%
\usepackage[title]{appendix}%
\usepackage{xcolor}%
\usepackage{textcomp}%
\usepackage{manyfoot}%
\usepackage{booktabs}%
\usepackage{algorithm}%
\usepackage{algorithmicx}%
\usepackage{algpseudocode}%
\usepackage{listings}%
\usepackage{arydshln}
\theoremstyle{thmstyleone}%
\theoremstyle{thmstyletwo}%

\theoremstyle{thmstylethree}%

\begin{document}


\title[Article Title]{Modelling handball outcomes using univariate and bivariate approaches} 

\author*[1]{\fnm{Dimitris} \sur{Karlis}}\email{ karlis@aueb.gr}

\author[2]{\fnm{Rouven} \sur{Michels}}\email{r.michels@uni-bielefeld.de}

\author[2]{\fnm{Marius} \sur{\"Otting}}\email{marius.oetting@uni-bielefeld.de}

\affil*[1]{\orgdiv{Department of Statistics}, \orgname{Athens University of Economics and
Business}, \orgaddress{\city{Athens}, \country{Greece}}}

\affil[2]{\orgdiv{Department of Business Administration and Economics}, \orgname{Bielefeld University}, \orgaddress{\city{Bielefeld}, \country{Germany}}}


\abstract{

Handball has received growing interest during the last years, including
academic research for many different aspects of the sport. On the other
hand modelling the outcome of the game has attracted less interest
mainly because of the additional challenges that occur. Data analysis
has revealed that the number of goals scored by each team are
under-dispersed relative to a Poisson distribution and hence new models
are needed for this purpose. Here we propose to circumvent the problem
by modelling the score difference. This removes the need for special
models since typical models for integer data like the Skellam
distribution can provide sufficient fit and thus reveal some of the
characteristics of the game. In the present paper  we propose some
models starting from a Skellam regression model and also considering
zero inflated versions as well as other discrete distributions in
$\mathbb Z$. Furthermore, we develop some bivariate models using copulas to model the
two halves of the game and thus providing insights on the game. Data
from German Bundesliga are used to show the potential of the new models.

}




\maketitle

\section{Introduction}
Handball is a dynamic sport, which is particularly popular in Europe and South America. In January 2024, the European Men's Handball Championship was held in Germany, where more than 50,000 spectators attended the opening match. Handball combines elements of football (soccer) and basketball, played on a rectangular court with two teams of seven players each. A handball match consists of a sequence of possessions, where each possession terminates with either a score or a loss of possession.

Handball is an invasion game, played for fixed time. Each match lasts for 60 minutes (two
halves of 30 minutes each), and is being played by two teams. 
The ball possession ends with a field shot attempt, or a seven meter throw, that is not rebounded by the offence (possession continues after an offensive rebound), or in case of
a turnover. 
The dynamics of the match process are revealed due to the difference of the players’ attack
and defence efficacy during the match, which influence the scoring rate of each team.

While there is an increasing steam of research on handball, models for final outcomes are less developed. Handball, contrary to football, is not a low-score sport and hence modelling its outcome can be challenging. For football, there are several proposals for modelling the score of a game (\citep{maher1982modelling, dixon1997modelling, michels2023extending}), mostly based on an assumption of a Poisson distribution for the number of goals. In the case of handball, 
a Poisson distribution for the number of goals for each team can be a starting assumption. Though, the number of goals often has a variance smaller than the mean making the Poisson assumption questionable. An alternative way is to assume a Gaussian distribution as an approximation of the Poisson distribution, thereby ignoring the discrete nature of the data. 

There are only a few existing models on handball scores. \citet{groll2020prediction} used a regularised Poisson regression model together with over- and underdispersed models. They also argue that since the Poisson mean is large enough, the model can be approximated by a standard Gaussian model (with sparsity also). 
\citet{smiatek2012statistical} analyse
team handball results of the German Bundesliga. Their
data set consists of ten seasons starting from 2001-02. While they do not have a particular model, the descriptive statistics indicate a small but consistent underdispersion. They also claim that the goal distribution can be described as a Gaussian distribution.
Recently, 
\citet{felice2023prediction}  
and \citet{felice2023ranking} proposed a Conway-Maxwell-Poisson distribution. This is a generalisation of the Poisson distribution that also allows for under-dispersion at the cost of added computational cost and rather approximating probability calculations. They also described some machine learning approaches to model the outcome and not the score. 
\citet{lago2013home} tried to model the goal difference using a simple regression model to examine how some technical characteristics affect the final goal difference but also to examine the home advantage. A similar approach has been used in \citet{prieto2016effects}.
In contrast to modelling match results, there is also some work on in-game modelling. \citet{dumangane2009departure} mention that
the dynamics of handball matches violate both the assumption of independence and identical distribution, in some cases having a non-stationary behaviour.
\citet{singh2023unified} extended this model.
They describe what is called the Markov-match, where the possession of the ball alternates between competitors and the probability of scoring, while different for each team, is quite high --- in fact, it can be up to $0.7$ for some teams. Thus, the number of goals cannot be considered as a binomial random variable (further approximated by a Poisson, if the goals are rare), since the individual Bernoulli trials (the attacks) are not independent. 



To model the winner of a handball match, we aim to overcome the problem of underdispersion by proposing alternative approaches. In particular, we 
consider the difference between the scores of the two competing teams. This difference, which is a number in $\mathbb Z= \{ \ldots,-2,-1,0,1,2,\ldots\}$, can be modelled as a realisation of the Skellam distribution. The Skellam distribution is a probability distribution that is frequently used in analysing the difference between two independent Poisson-distributed random variables. In this paper, we use the Skellam distribution to model the final score difference of handball matches. However, as the final outcomes of matches exhibit special patterns, we aim to explicitly address these in two different ways. First, we consider a zero-inflated (ZI) Skellam distribution which shifts probabilities to the zero from the other values, as well as other discretised distributions defined in $\mathbb Z$ . Second, we investigate the score difference of the first and second half separately as those differences do \textit{not} indicate systematic deviances. We model these differences jointly by constructing bivariate distributions using copulas. Modelling the scores of the first and second halves in such a way enables us to calculate conditional probabilities of the final score based on the score at half time. Through this approach, we aim to contribute to the evolving landscape of sports analytics, providing a useful framework for understanding and predicting the outcomes in handball.

The rest of the paper is structured as follows. Section~2 introduces the univariate and bivariate distributions we use. These are applied in Section~3, which presents models fitted to handball scores from the Bundesliga in Germany. Concluding remarks can be found in Section~4.




\section{Methods}

This section first introduces the Skellam distribution, which is then extended to obtain the zero-inflated Skellam distribution. Some other candidate models like the discretized versions of normal and Laplace distributions are also described.  At the end of the section, we consider copula-based models with Skellam marginals.

\subsection{Skellam distribution}

In handball, the score differences are represented as negative and positive integers, that is values in $\mathbb Z$.
Perhaps the most well-known distribution in $\mathbb Z$ is the Skellam distribution. 
\citet{irwin1937frequency} derived the distribution of the difference of two independent variables following the Poisson distribution with the same
mean, and \citet{jg1946frequency} generalised this to different means. 

If the random variables  $X$ and $Y$ follow independent Poisson distributions with
parameters $\theta_1>0$ and $\theta_2>0$, then the
random variable $Z=X-Y$ has probability mass function (pmf) given by \begin{equation}
     P(Z=z|\theta_1, \theta_2)
                             =e^{ -(\theta_1+\theta_2) }
                              \left( \frac{\theta_1}{\theta_2} \right)^{z/2}
                               I_{|z|}\left(2 \sqrt{\theta_1 \theta_2}\right), ~~z \in \mathbb Z,~~\theta_1,\theta_2>0,
\label {diff}  \end{equation} 
where $I_{r}(x)$ is the modified Bessel function of order $r$
\citep[][pp. 375]{mABR70a} 
 \[ I_r(x)=\left(
\frac{x}{2} \right)^r
       \sum \limits _{m=0}^{\infty}\ \frac{\left( \frac{x^2}{4} \right)^m}
                                          { m!\Gamma(r+m+1)}.
\]
We will denote this distribution as the Skellam$(\theta_1,\theta_2)$ distribution. The Skellam distribution has also been called the Poisson difference distribution. 
Its mean and variance are ${\mathrm E}(Z)=\theta_1-\theta_2$ and
$\mbox{Var}(Z)=\theta_1+\theta_2$. This can be used  to reparameterize the distribution; clearly, $\mbox{Var}(Z) \ge
|{\mathrm E}(Z)|$.
The skewness is given as $(\theta_1-\theta_2)/(\theta_1+\theta_2)^{3/2}$.
and it is determined by the sign of $\theta_1-\theta_2$; in the case of $\theta_1=\theta_2$, we come up with the symmetric Skellam distribution.  
For large values of $\theta_1+\theta_2$,
the distribution can be well approximated by the normal
distribution. If $\theta_2$ is very close to zero, then the
distribution tends to a Poisson distribution. If the parameter $\theta_1$ approaches zero, then the distribution is the negative of
a Poisson distribution. The Skellam distribution is unimodal.

We can derive the
Skellam distribution as the difference of other distributions as
well \citep{KarlisNtzoufras06SIM}, which motivates its use in various applications. 
In particular, it 
can be derived as the difference of two underdispersed random variables as follows:  Let $X_1$ and $X_2$ be two Poisson random variables and $W$ be an underdispersed random variable, i.e.\ with variance less that the mean, independently of $X_1$ and $X_2$. Consider the random variables $Y_1 = X_1 +W$ and $Y_2=X_2 + W$. Then, 
$Y_1$ and $Y_2$ will be underdispersed. If we take their difference, we obtain $Y_1-Y_2 = X_1+W -X_2 -W = X_1-X_2$ which is a Skellam random variable being the difference of two Poisson variables. This makes the assumed Skellam model as a reasonable approach for underdispersed data, such as the data on handball considered in this work. Note that we do not need to assume some parametric assumption for the two random variables, just for their difference, as in the above definition $W$ can follow any underdispersed distribution. 

To allow for better interpretation of the parameters but also more advanced modelling approaches such as regression, \citet{koopman2017intraday} use a reparametrised version of the distribution with mean $\mu =\theta_1-\theta_2$
and variance $\sigma^2 = \theta_1 +\theta_2$. 
We will denote this by Skellam2$(\mu,\sigma^2)$ in the following. 
Here, one can add covariates in
the mean as a simple linear model since the mean is defined in $\mathbb R$, see \citet{pelechrinis2021skellam, holmes2024forecasting}. 
%
To cope with data for which the regular Skellam distribution lacks a proper fit, the baseline model has been extended in various ways. \citet{KarlisNtzoufras06SIM} consider a zero-inflated version, while a zero-deflated version is given in \citet{koopman2017intraday}. On a similar note,
\citet{ntzoufras2021bayesian} propose a truncated version of the Skellam distribution.
A mixture of Skellam distributions has been used by \citet{jiang2014skellam} to cluster differentially expressed genes.  
A modified version where probability mass is exchanged between certain values is given in \citet{koopman2017intraday}. 
For a recent review on the Skellam distribution see \citet{tomy2022retrospective,karlis2023models}.

\subsection{Zero-inflated Skellam distribution}
From the extensions summarised above, we consider the zero-inflated version in the following, since draws are overrepresented the handball data (corresponding to a score difference of zero).  
The pmf of the zero-inflated Skellam distribution is given by
\[
P_{Z}(Z=z|\theta_1,\theta_2,p) =
\left\{
\begin{array}{cc}
p + (1-p) P(Z=z|\theta_1,\theta_2) & \mbox{if}~~z=0, \\
(1-p) P(Z=z|\theta_1,\theta_2)&
\mbox{if}~~z \ne 0. 
\end{array}
\right.
\]
with $p \in [0,1)$ and $P(\cdot)$ being the pmf of the Skellam distribution given in (\ref{diff}). Thus, the zero-inflated Skellam distribution shifts some probability from the other values towards  zero.

The model can be fitted by the EM algorithm, which is a standard choice for fitting zero inflated models. At first, we are not using covariates for modelling the parameter $p$. We thus make a rather unnatural assumption that each match, despite the difference between the teams' abilities, has a constant probability to end in a draw. However, we believe that this assumption makes sense to avoid overparametrising the model. 

A model that allows for covariates on the inflation parameter is also possible by assuming that for match $i$
\[
p_i = \frac{\exp({\bf z}_i' \boldsymbol{\gamma})}{1+ \exp({\bf z}_i' \boldsymbol{\gamma})}
\]
where ${\bf z}_i$ is a vector with explanatory variables related to match $i$ and $\boldsymbol{\gamma}$ a coefficient vector to be estimated. 

\subsection{Other univariate distributions in $\mathbb Z$}
Alternative models for modelling data in $\mathbb Z$ can be also considered either as the difference of discrete random variables or by discretizing continuous ones (for an overview of such models see \cite{karlis2023models}). For example, a discrete normal distribution can be considered based on a $\mathcal{N}(\mu,\sigma^2)$ distribution as
\begin{eqnarray*}
P_N(Z=z| \mu, \sigma^2) &=&  \int_{z-0.5}^{z+0.5} \phi(t; \mu, \sigma^2) dt \\
&=&\Phi(z+0.5; \mu,\sigma^2) - \Phi(z-0.5; \mu,\sigma^2)
\label{discretenormal}
\end{eqnarray*}
where $\phi(z; \mu,\sigma^2)$ and  $\Phi(z; \mu,\sigma^2)$
are the density and  the cumulative distribution function of a $\mathcal{N}(\mu,\sigma^2)$, respectively. 
An important remark, which is also a motivation for our approach, is that the Skellam allows for skewness which is not the case for the discrete normal. 

Here, the same parameterisation as for the Skellam distribution is used for $\mu$. 
In particular, $\mu_i$ is related to some covariates that reflect the ability of the teams. Note that we here consider the discrete normal which is a different approach than fitting a standard regression and then to simply discretise the results of the regression.

Another option is to use the discretised version of the Laplace distribution \citep{barbiero2014alternative}. 
There are different ways to create a discretised version of the Laplace, here we use the simple idea, as for the normal above, to take the cumulative function of the continuous case at specific points. Thus, we have a pmf defined as
\begin{equation}
P_L(Z=z| \mu, \sigma^2) =  F(z+0.5; \mu,\sigma^2) - F(z-0.5; \mu,\sigma^2)
\label{discreteLaplace}
\end{equation}
where $F(x;\mu,\sigma^2)$ is the cumulative distribution of a Laplace distribution with location $\mu$ and scale $\sigma^2$. Such a model provides more mass at zero, recall the shape of the Laplace distribution. 


\subsection{Bivariate distributions with copulas}
\begin{figure}
\begin{center}
\includegraphics[scale=0.50]{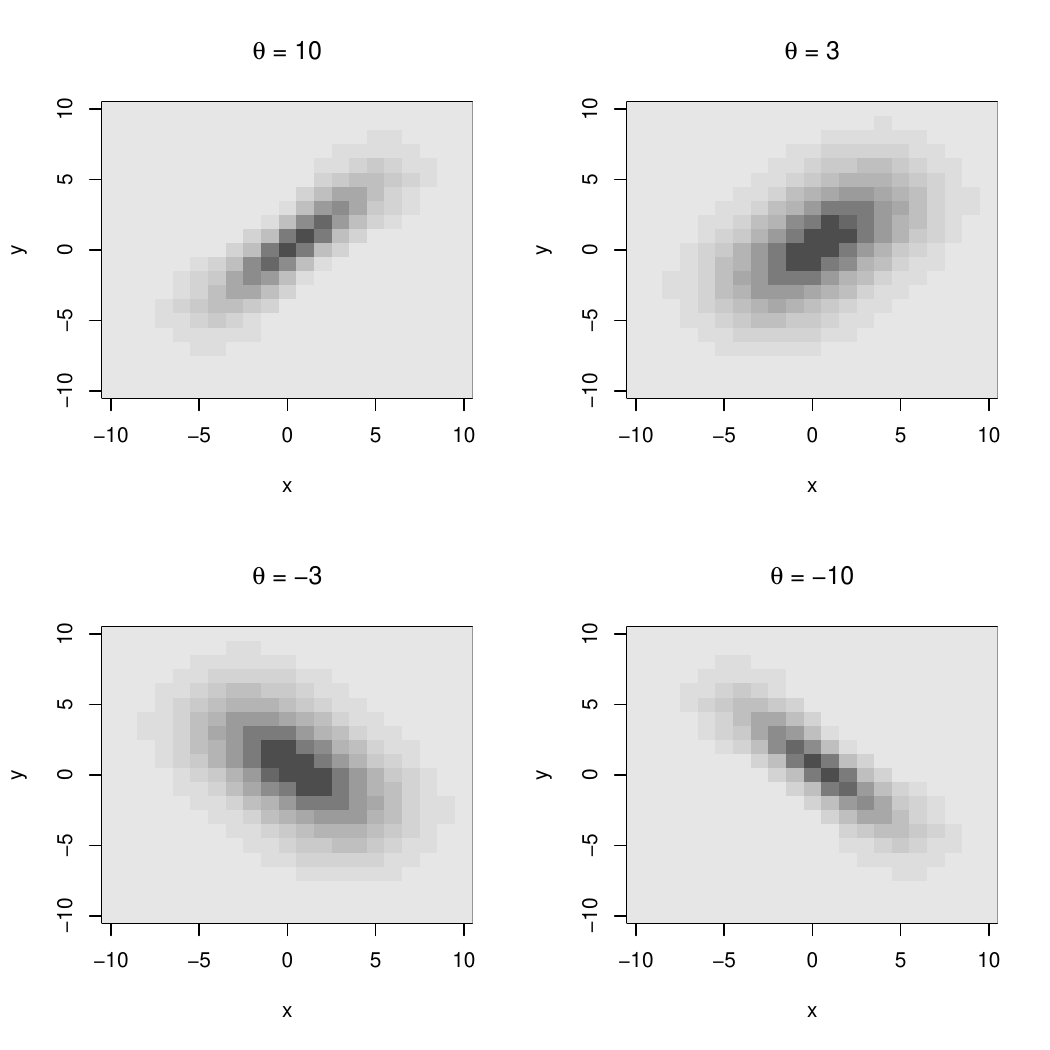}
\caption{ \label{skel1} Marginal distributions are Skellam
$\mu_1=0.5$ $\mu_2=0.5$, $\sigma_1^2=15$ , $\sigma_2^2=14$, which are reasonable values for the case study of handball scores, and varying
copula parameter for the Frank copula. The dependence parameter values are 10, 3, -3
and -10. 
}
\end{center}
\end{figure}

To provide a more thorough analysis of the results of handball matches, the score difference can also be modelled for each half separately. 
In this case, denote as $Y_1$ and $Y_2$ the score difference for the first and second half.
To jointly model such score differences, we need a bivariate distribution defined in ${\mathbb Z}^2$.
We can construct such a distribution accounting for dependencies between both halves using copulas. 
In particular, we assume that marginally both of the $Y_1$ and $Y_2$ follow a Skellam distribution with parameters
$\mu_j,\sigma_j^2, \ j = 1,2$, coupled with some copula $C(\cdot,\cdot;\theta)$, where $\theta$ is the
dependence parameter.

For the discrete case, we need to
take differences to derive the joint pmf. For example, the
bivariate case with marginals $F(y_1)$ and $G(y_2)$ one can derive the
joint pmf as
\begin{eqnarray*}
P(Y_1=y_1, Y_2=y_2)&=& C\big(F(y_1),G(y_2);\theta\big)-C\big(F(y_1-1),G(y_2);\theta\big)- \\
&& C\big(F(y_1),G(y_2-1);\theta\big)+C\big(F(y_1-1),G(y_2-1);\theta\big)
\end{eqnarray*}

We are going to use Skellam marginal distributions
linked through the bivariate Frank copula, which is given by
\[
C(u,v; \theta)=-\frac{1}{\theta}\log{\left[1+\frac{(\exp^{-\theta
u}-1)(\exp^{-\theta v}-1)}{(\exp^{-\theta}-1)}\right]}
\]
and a Gumbel copula with cdf
\[
C(u,v; \theta) = \exp\left(-((-\log(u))^\theta + (-\log(v))^\theta)^{1/\theta} \right).
\]

We select the two copulas mainly to be as flexible as necessary since it allows for both negative and
positive correlation (Frank copula) and for tail dependence (Gumbel copula). In our case, tail dependence means that 
for larger differences we expect to occur both in the first and second half together, implying also that they can occur more often than a model without (or small) tail dependence.  

\begin{figure}
\begin{center}
\includegraphics[scale=0.50]{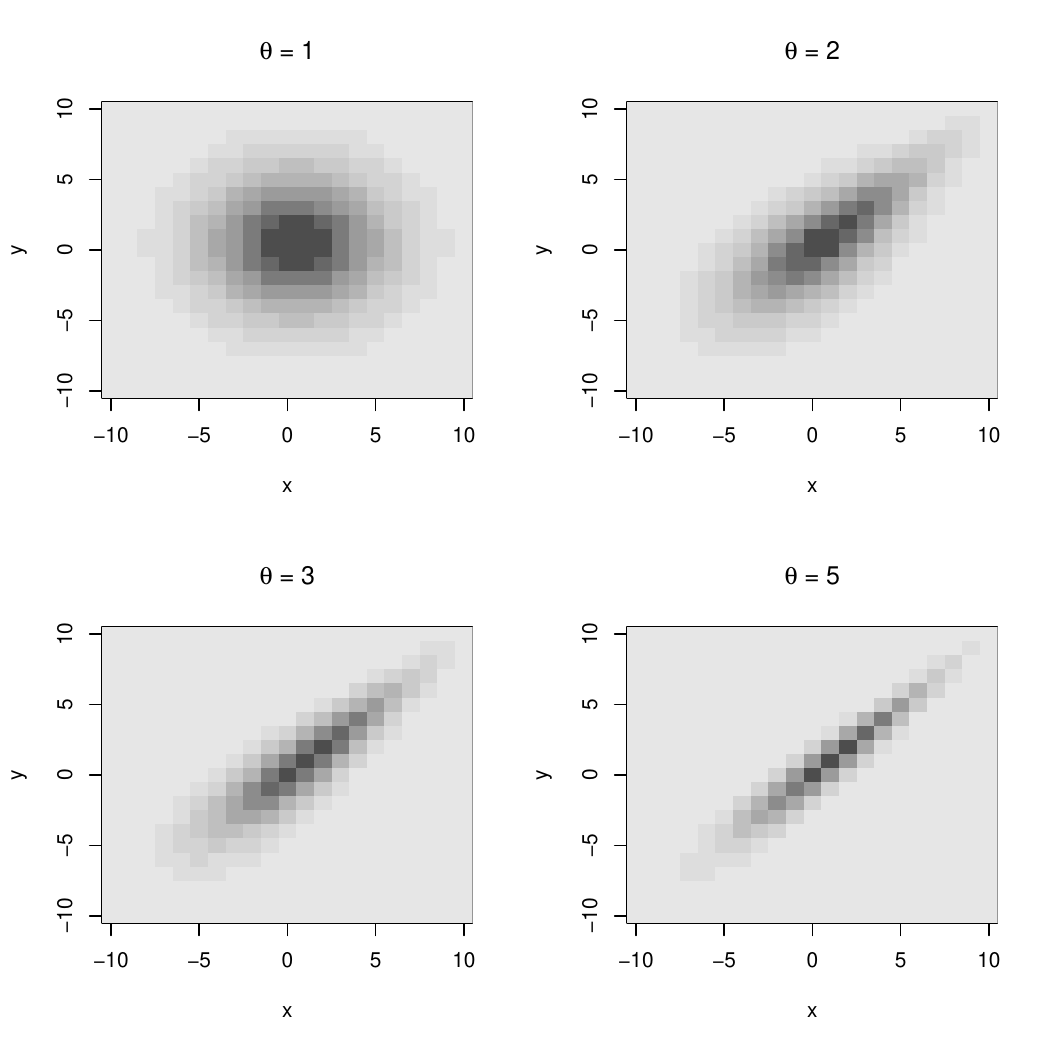}
\caption{ \label{skel2} Marginal distributions are Skellam
$\mu_1=0.5$ $\mu_2=0.5$, $\sigma_1^2=15$ , $\sigma_2^2=14$ and varying
copula parameter for the Gumbel copula. The values are 1 (independence),2,3 and 5. 
}
\end{center}
\end{figure}

However, of course any other copula could also be used. The approach further applies for any marginal distribution in $\mathbb Z$.
Marginal properties of the model are based on the univariate Skellam.
Figures  \ref{skel1} and  \ref{skel2} depict the joint probability
function with Skellam marginals and a Frank and Gumbel copula respectively with various
parameters. We have used marginal distribution close to what we observe in handball and varying dependence structure including both positive and negative dependence. 


For a broad review on bivariate Skellam and related distributions, see \citet{karlis2023models}.


\section{Application to German Handball Bundesliga}
In this section, we first introduce the data considered to fit the models presented in the previous section. Afterwards, we report the estimation results and subsequent analyses for the univariate and bivariate distributions.

\subsection{Data}
Our data consist of the scores in the German Handball Bundesliga (HBL) for the seasons 2017-18 up to 2022-23
which comprise, in total, 1844 matches.
The HBL is the top German professional handball league with 18 teams competing in a round-robin format. 
The bottom two teams are relegated to the 2$^{\text{nd}}$  Bundesliga, while the top two teams of the 2$^{\text{nd}}$  Bundesliga will be directly promoted to the Bundesliga.

\begin{center}
\begin{figure}
\centering
\includegraphics[width = 0.7\textwidth]{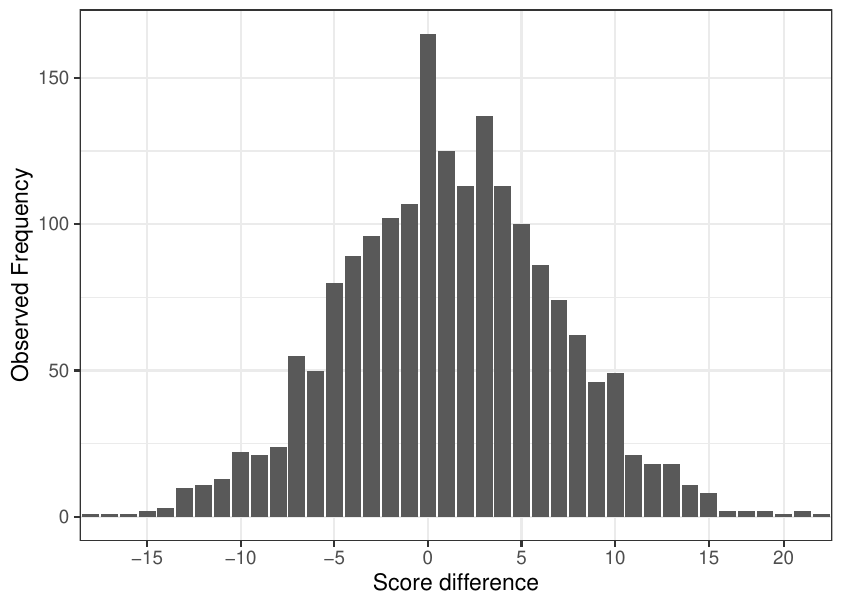}
\caption{\label{hist} Observed frequencies of the score differences in the season 2017-18-2022-23.}
\end{figure}
\end{center}

The observed means and variances for the home goals are $28.07$ and $19.67$, and $26.92$ and $18.32$ for the away goals, thus implying 
a small home advantage, but also underdispersion.
For an attempt to model such underdispersion see
\citet{felice2023ranking}, who uses a COM Poisson distribution fitted to the home and away goals.
In addition, the observed correlation between the home and away goals in our data is relatively small (0.14). This is somewhat counter-intuitive, since in handball the two teams
are in offense sequentially, hence one could expect to observe a higher correlation between the number of goals scored. 
To investigate whether such patterns change over time, and also to test the stability and robustness of the models, we investigate the seasons considered in our analysis separately. 
Table \ref{descr} displays the season-by-season average goal difference per game, their standard deviations as well as 
the skewness of the goal difference. 
The latter can be a motivation for considering the Skellam model as we observe some skewness. Note that season 2019-20 was cancelled after 240 matches due to COVID-19 and, consequently, 20 teams participated in the 2020-2021 season instead of 18 (in total 380 matches). For both seasons, we can observe a drop in the mean which corresponds to a smaller margin of home wins on average. This coincides with the well-known finding that the home advantage was mitigated during games behind closed doors (\cite{bryson2021causal}).

\begin{table}[ht]
\centering
\begin{tabular}{lccc}
 Season & Mean & SD & Skewness \\ 
  \hline
2017-18 & 1.28 & 5.83 & -0.06 \\ 
2018-19 & 1.21 & 5.75 & 0.04 \\ 
2019-20 & 1.33 & 5.04 & 0.03 \\ 
2020-21 & 1.07 & 5.65 & 0.19 \\ 
2021-22 & 0.71 & 5.90 & 0.04 \\ 
2022-23 & 1.35 & 5.96 & -0.10 \\ 
   \hline
\end{tabular}
\caption{\label{descr} Descriptive statistics for the score differences in the six years of data.}
\end{table}

\subsection{Univariate analysis}

In this section, we consider the mean/variance parametrization of the Skellam2 model such that we are able to add covariates.
We model the mean goal difference $\mu_{jk}$ between the two teams, i.e.\
\begin{eqnarray*}
\mu_{jk} &=& \alpha + \beta_j + \gamma_k,
\end{eqnarray*}
where team $j$ plays at home against team $k$. In this model formulation, $\beta_j$ is the ability of the $j$-th team at home while $\gamma_j$ is the ability away. The team ``Bergischer HC'' has been used as the baseline team. So the interpretation of $\beta_j$ is that when team $j$ plays at home with Bergischer the expected score is $\alpha+\beta_j$ while if they play away (so at home of Bergischer) it is $\alpha+\gamma_j$. Their difference $\beta_j -\gamma_j$ measures the home advantage for the $j$-th team. The model explicitly assumes a different home advantage for each team.  
 Home advantage has been discussed for handball in many instances, see for example the work in \cite{volossovitch2021home}.
We employ the same parametrisation of $\mu_{jk}$ also for the zero-inflated version as well as for the discrete normal and discrete Laplace distributions.

We have considered a regular Skellam model, followed by the zero-inflated version. The latter has an extra additional parameter $p$ that reflects the extra zeros we have observed. In the case of handball, the idea behind zero inflation is that teams that are close to the score at the end of the game do not want to risk and prefer the safe path of a draw. Recall that draws are not common in handball as they are in football but it is still a possible outcome. 
Moreover, we try other discrete distributions as well, such as the discrete normal or the Laplace distribution.
Finally we have also considered models with covariates in $p$, i.e. the zero inflation parameter.
One reason for that is that using a common $p$ for all matches is somewhat unrealistic. An interpretation of zero inflated models is some additional zeros, i.e. draws in our case, that cannot be predicted by the model.  It seems irrelevant to assume this when we have teams of very different strength to be the same as for the case of equal strength teams. For this reason some covariate may differentiate this inflation probability. We have used as a covariate the betting odds at the beginning of the game as an indication of the strength difference of the two teams. The results did not show any statistical significance and the log-likelihood improved very little. For this reason we skip them. 

\subsubsection{Model fit}
\begin{table}[ht]
\centering
\begin{tabular}{lcccccc}
  \hline
 & 2017-18 & 2018-19 & 2019-20 & 2020-21 & 2021-22 & 2022-23 \\ 
  \hline
Skellam & 1755.78 & \textbf{1743.45} & 1387.50 & 2245.23 & 1854.39 & \textbf{1813.42} \\ 
  ZI Skellam & \textbf{1751.07} & 1745.45 & 1389.50 & \textbf{2244.96} & \textbf{1850.77 }& 1815.42 \\ 
  Discrete normal & 1756.25 & 1743.92 & 1386.47 & 2245.31 & 1855.83 & 1814.38 \\ 
  Discrete Laplace & 1765.62 & 1758.51 & \textbf{1376.45} & 2253.07 & 1854.42 & 1820.88 \\
   \hline
\end{tabular}
\caption{\label{results} AIC results from fitting several models to the data from different periods as well as the resulting ZI parameter in the ZI Skellam model.}
\end{table}

We have fitted the Skellam model, the zero-inflated Skellam, the discrete normal and the discrete Laplace. When fitting these models to the data, we consider a season-by-season approach to check whether the discrepancy seen in Figure~\ref{hist} is not primarily driven by a single season.
Table \ref{results} displays AIC values for the models in the different seasons. The zero-inflated model provides slightly better log-likelihood at the cost of an additional parameter, so judging from AIC it is not a preferable model in every season. The discrete normal model provides slightly worst fit due to the small skewness present in our data. Note that in 2019-20  season we observe a restricted number of observations due to the cancelled seasons, which is the reason why the structure of that season is least reliable. In all the other seasons, either the Skellam or the zero-inflated Skellam are chosen, while the latter one automatically includes the general Skellam distribution which is why from a log-likelihood point of view, the zero-inflated Skellam is chosen for every full played season.

\subsubsection{Goodness-of-fit}
In this subsection, we evaluate the goodness-of-fit. For this, we consider the first season that was held without any COVID-19 restrictions, i.e.\ the 2021-22 season. 


\begin{table}[ht]
\centering
\begin{tabular}{l|rrr}
 & home wins & draw & away wins \\ 
  \hline
  \hline
Observed & 153.00 & 29.00 & 124.00 \\ 
\hline
  Bets implied & 160.63 & 28.22 & 117.14 \\ 
  \hdashline
  Skellam & 157.45 & 20.66 & 127.89 \\ 
  ZI Skellam & 152.02 & 30.39 & 123.59 \\ 
  Discrete Normal & 157.55 & 20.40 & 128.05 \\ 
  Discrete Laplace & 159.95 & 21.35 & 124.70 \\ 
   \hline
\end{tabular}
\caption{\label{odds} Expected outcomes over the 306 matches of the 2021-22 season based on different models and bookmaker's odds.}
\end{table}

To evaluate the goodness-of-fit of the models, we proceed as follows: 
Based on the fitted model, we predict for each match the probabilities of a home win, a draw and an away win. We sum these predicted probabilities over all matches and compare them to the corresponding actual frequencies found in the data. We further consider betting odds\footnote{Betting odds were taken from
\url{https://www.betexplorer.com/handball/germany/bundesliga-2017-2018/results/}}, which were transformed to probabilities by assuming an equal vig for the three outcomes. 
Table \ref{odds} shows the results. The baseline Skellam model predicts ten fewer draws than we observe in the data, which is not negligible.
Thus, we can summarise that the fit of the Skellam is sufficient but it has some failure for the draws, which can be corrected with the zero-inflated Skellam model. The discrete normal is somewhat inferior as we have seen. This is confirmed when comparing the models to the predictive performance of bookmakers: here, the zero-inflated Skellam can not only keep up with the bookmaker's accuracy for draws (which is kind of the targeting outcome for the ZI Skellam) but is also able to predict home- and away wins more precisely than bookmakers. This is particularly astonishing considering that bookmakers are able to include short-term news such as injuries of important players into their odds which we did not include in the ZI Skellam.

\begin{center}
\begin{figure}
\centering
\includegraphics[scale=0.5]{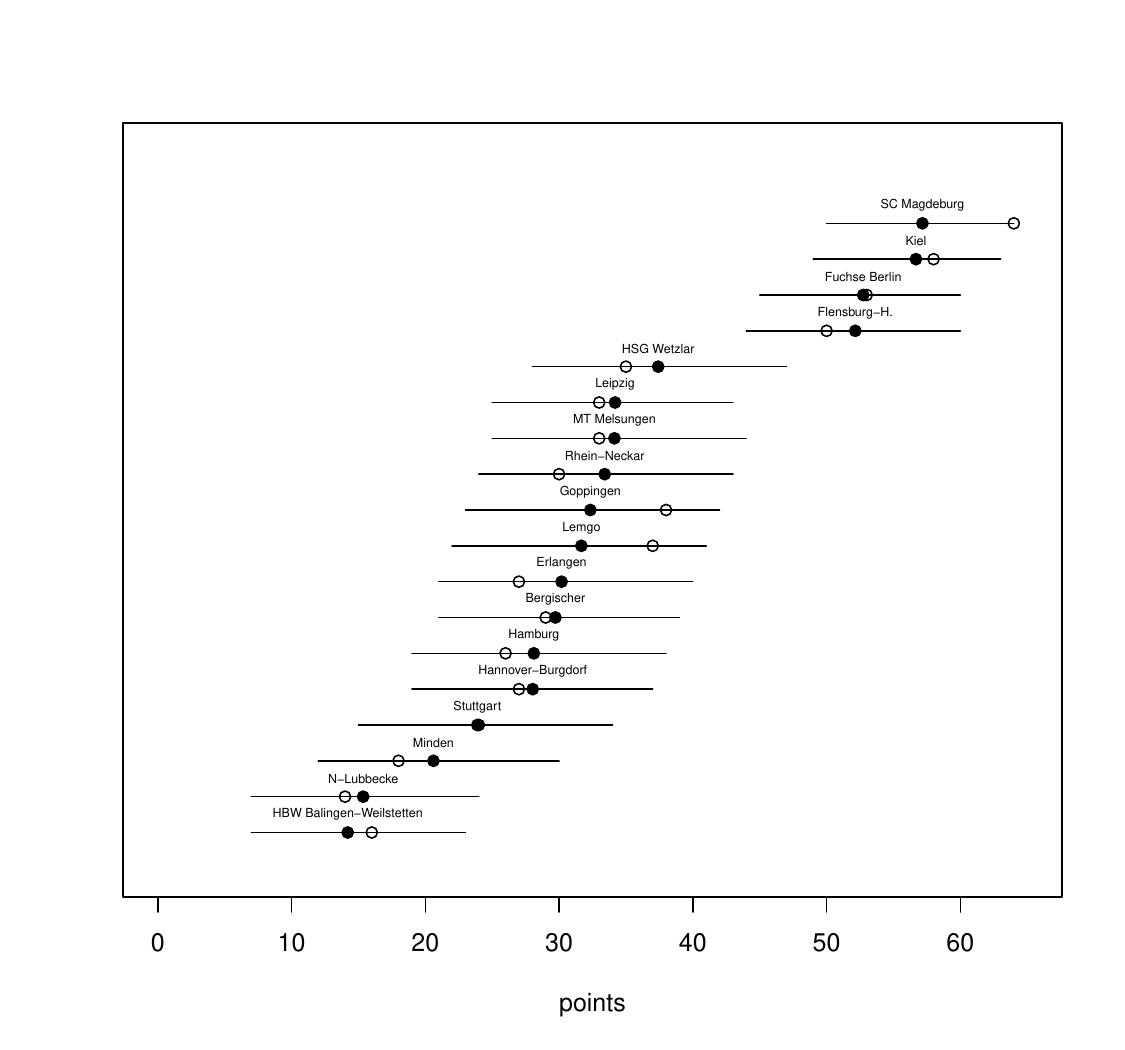}
\caption{\label{champ2} The observed numbers of points (blank points), the expected number of points through simulations (filled points) and 95\% confidence intervals around.}
\end{figure}
\end{center}

As we have now seen that the zero-inflated Skellam model is superior in terms of the AIC in the most seasons and the prediction of draws in the 2021-22 season, we perform the remaining goodness-of-fit checks using this model.

Figure \ref{champ2} shows the observed number of points for each team, the expected number of points based on 10,000 simulations of the league and the 95\% confidence interval as obtained via parametric bootstrap and use of the percentile method. According to our results, SC Magdeburg does not seem to be a clear champion, since Kiel also had also reasonable chances to win the championship --- however, SC Magdeburg won the championship by a margin of 6 points.

\begin{center}
\begin{figure}
\centering
\includegraphics[scale=0.4]{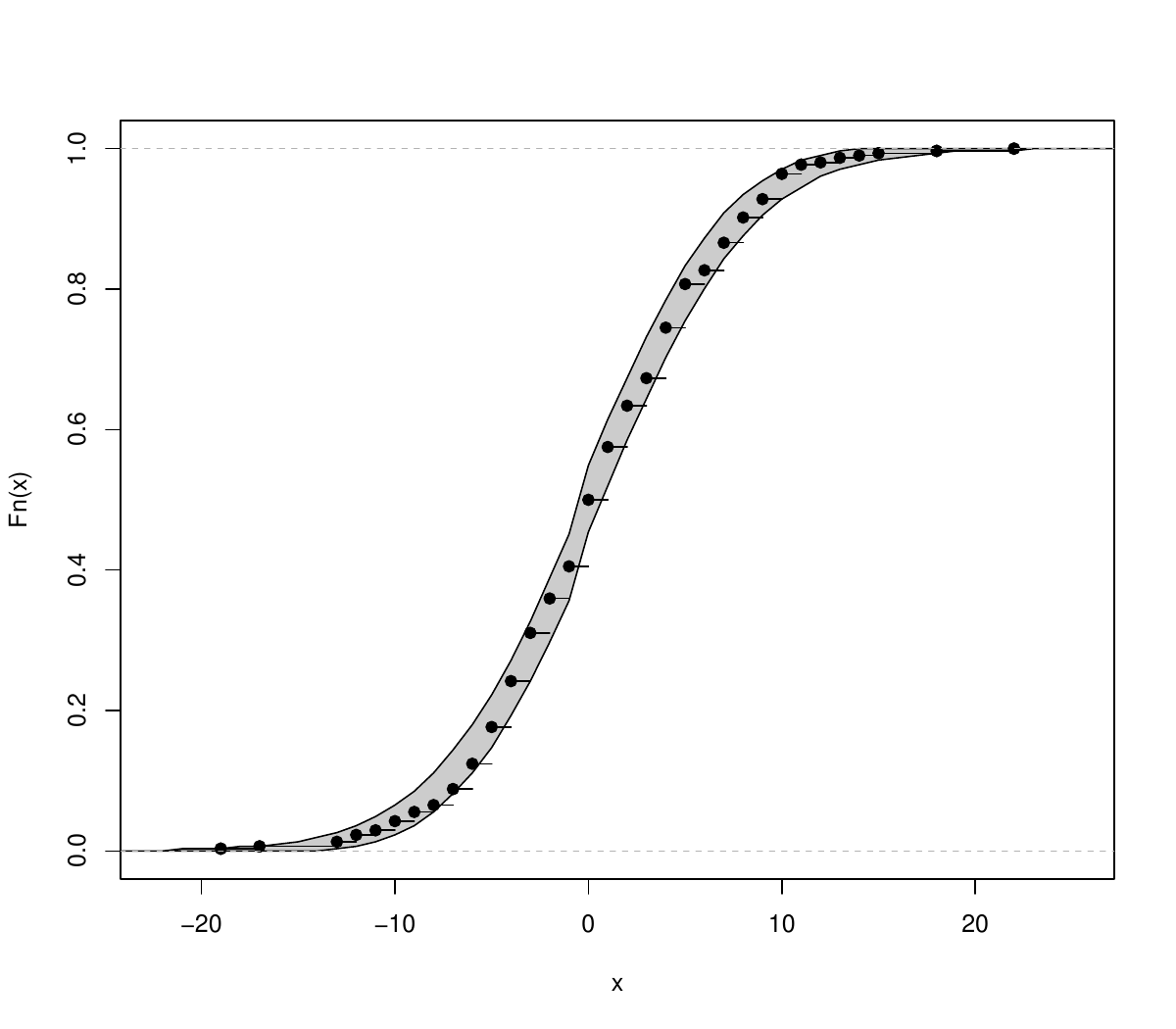}
\caption{\label{gof} Empirical distribution function of the observed score differences and a 95\% confidence interval under the ZI Skellam model.}
\end{figure}
\end{center}

As another goodness-of-fit check, we again consider the 10,000 simulated championship results and investigate the corresponding empirical distribution function. 
Based on all the replications, we estimated a 95\% confidence interval for $P(Y \le y)$, which is indicated by the gray area in Figure \ref{gof}. The actual empirical cumulative distribution function (cdf) is indicated by the black dots, demonstrating that there is only a minor deviation compared to what we expect under the fitted model. 

\subsubsection{Out-of-sample prediction}

\begin{figure}
\centering
\includegraphics[scale=0.5]{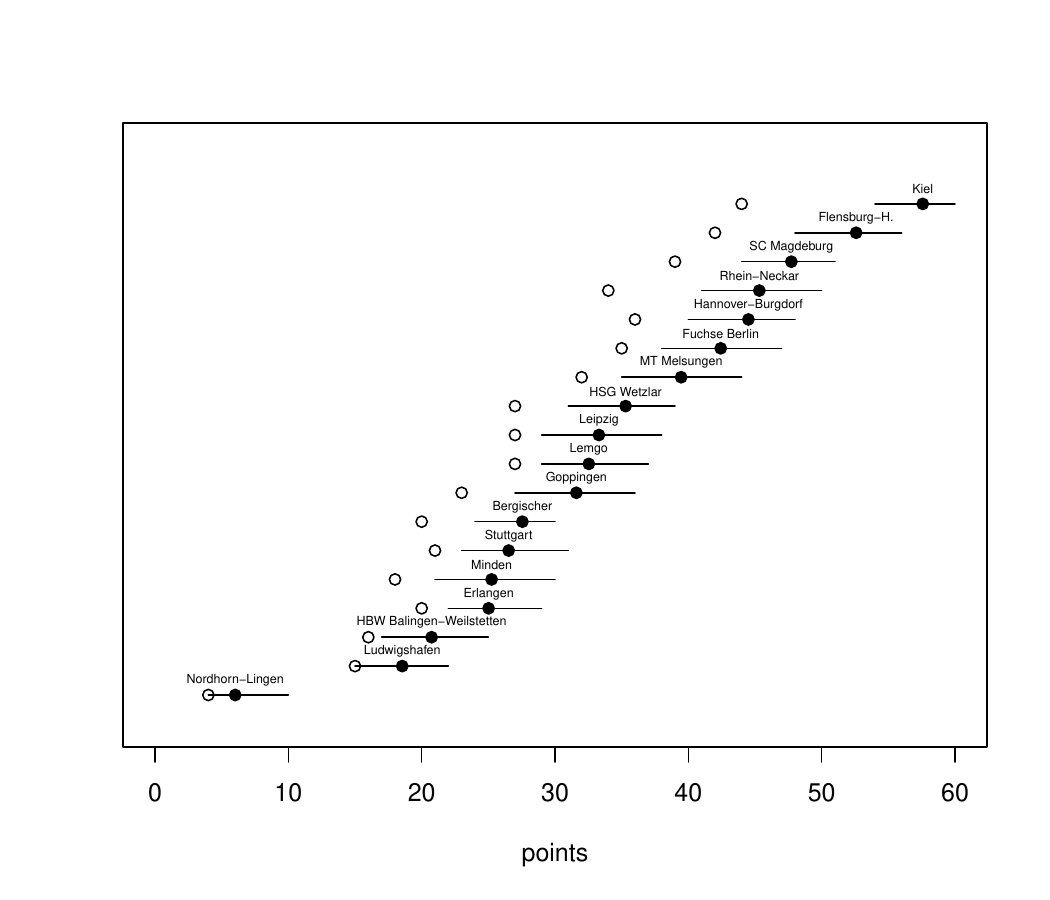}
\caption{\label{champ3} The expected points (filled) together with 95\% confidence intervals in the COVID-19 season if the season would have been finished under the prediction using the zero-inflated Skellam. The blank points indicate the number of points when the season was interrupted.}
\end{figure}

In this section, we present an analysis showcasing the feasibility of out-of-sample prediction, using the context of the disrupted COVID-19 season as an illustrative case study. Following the approach outlined by \citet{csato2021coronavirus} and \citet{van2023probabilistic}, we employ a Monte Carlo simulation comprising 10,000 iterations to compute the probabilistic end-of-season standings under the assumption of a completed regular season. 
The model was fitted to the data available until the stop and then the estmated parameters were used for the rest of the season. 
Figure \ref{champ3} depicts the hypothetical end-of-season table, while Figure \ref{champ2} illustrates the expected remaining points to be accrued. Notably, despite a significant number of fixtures remaining unplayed, the expected season winner aligns with the outcome derived from terminating the season prematurely. This observation is congruent with the findings presented in Table \ref{champtab}, which displays the likelihood of winning the league  and being relegated following the season's interruption, as determined by the Monte Carlo simulation.

Furthermore, our analysis extends to examining the hypothetical relegation scenarios (notwithstanding the absence of relegation in the respective year). Of particular interest is the favourable outcome for Nordhorn-Lingen, who consistently faced relegation across all simulation iterations, but were not relegated at the end.

Moreover, the league administration faced the task of determining eligibility for participation in international competitions in the subsequent year. Initially, all teams up to Fuchse Berlin received international tickets, presenting a unique circumstance for MT Melsungen, whose projected point came close to the ones of Fuchse Berlin. However, an opportune withdrawal by Hannover-Burgdorf facilitated MT Melsungen's ascension to an international slot and, thus prevented discussions about the fairness of the league's decision.

\begin{table}[ht]
\centering
\begin{tabular}{lcc}
  \hline
 team & Champion & Relegation \\ 
  \hline
Kiel & 0.972 & 0.000 \\ 
Flensburg-H. & 0.028 & 0.000 \\ 
Stuttgart & 0.000 & 0.001 \\ 
Erlangen & 0.000 & 0.002 \\ 
Minden & 0.000 & 0.006 \\ 
HBW Balingen-Weilstetten & 0.000 & 0.273 \\ 
Ludwigshafen & 0.000 & 0.719 \\ 
Nordhorn-Lingen & 0.000 & 1.000 \\ 
   \hline
\end{tabular}
\caption{\label{champtab} Probabilities for winning the championship/being relegated if the 2019-20 season would have been continued. For all other teams both probabilities were zero.}
\end{table}

\subsection{Bivariate analysis}
In sports analytics, copulas have been used to model jointly the number of
goals scored in soccer \citep[see, e.g.][]{mchale2007modelling}.
Here we apply them for modelling the difference in the number of goals between the two halves.

An interesting result about the Skellam distribution is an additive property:
the sum of Skellam random variables is also a Skellam. In a similar manner,
for the Skellam distribution we have that if a random variable $Y$ follows a Skellam in the unit of time, then it follows a Skellam for any other unit of time. In our bivariate analysis this simply implies that we may assume that each half of the match the score difference also follows a Skellam distribution. In the following, we investigate whether the scoring abilities remain the same across the two halves of the game. 

\begin{center}
\begin{figure}
\centering
\includegraphics[scale=0.6]{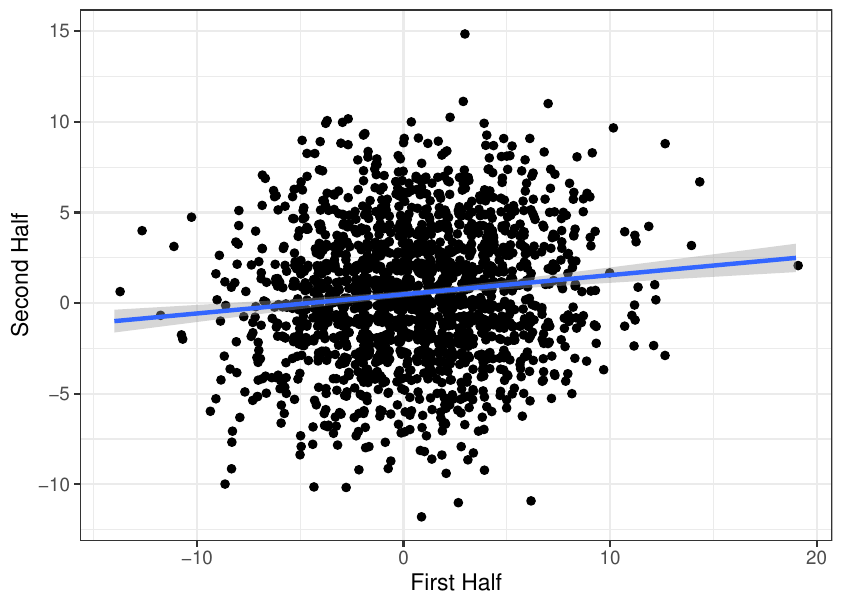}
\caption{\label{cond1}  Correlation between the score differences in the two separated halves for all seasons. We do observe a slightly positive correlation (0.13), rendering the Frank copula as suited.}
\end{figure}
\end{center}

\begin{center}
\begin{figure}
\centering
\includegraphics[scale=0.7]{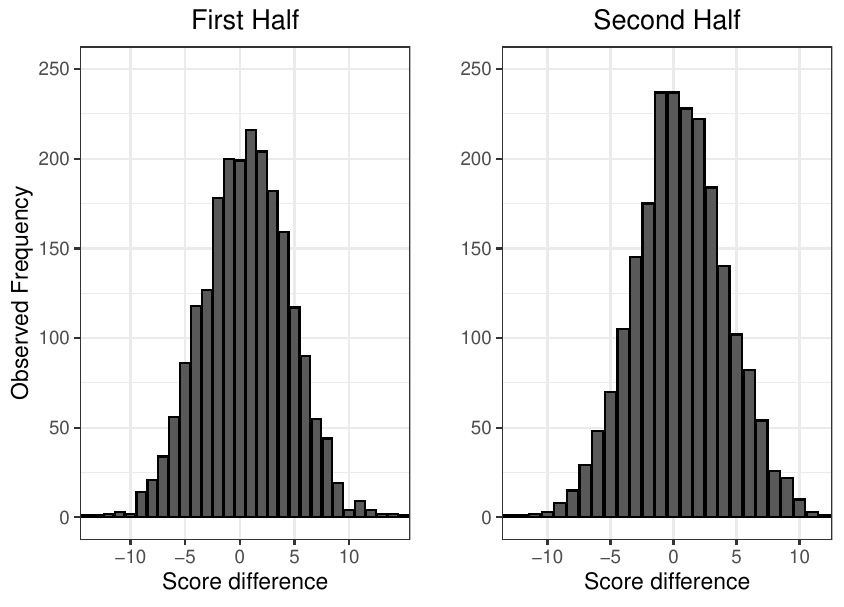}
\caption{\label{cond2} Histogram of the score differences in the two separated halves for all seasons. We do not observe any structural errors, thus, the Skellam seems to be suited as the marginal distribution for the Copula. }
\end{figure}
\end{center}

We consider the data from the German Bundesliga 2021-22 to model the score differences in the first and second half. For the copulas, we consider a Frank and a Gumbel copula.

As for the univariate case, we extend the basic bivariate model to include covariates. Again, we work with
two random variables $Y_{j1}$ and $Y_{j2}$, where each follows marginally a Skellam2 distribution with mean $\mu_j$ and variance
$\sigma_j^2$, $j=1,2.$ Mean parameters are related to some covariate vector ${\bf
Z}_{ji} $ with $p$ elements. Clearly each variable can be related
to a different set of covariates. We also assume a linear
relationship,
\begin{eqnarray*}
\mu_{1i} &=& \beta_1^T {\bf Z}_{1i} \\
\mu_{2i} &=& \beta_2^T {\bf Z}_{2i}, \\
\end{eqnarray*}
where $\beta_j$ are the regression coefficient vectors $j=1,2$. We
further model any dependence between the two variables using a copula (Frank and Gumbel in our case)
with parameter $\theta$, which may also depend on
covariates. Based on this parametrization, we need to estimate
$\Theta= (\beta_1,\beta_2,\sigma_1^2,\sigma_2^2, \theta)$. 

Such bivariate models offer a great flexibility for modelling handball results:
we can, for example, estimate the probability that a team will win, the score difference at the end of the match or even conditional probabilities that a team wins the match given a score difference at half time.

We have fitted different models, thereby gradually increasing the model's complexity.
Model A assumes independence, i.e.\ the two variables are fitted separately without a copula. For each one we assume different effects, so we assume that the abilities of the teams are different. 
Model B assumes dependence through a Frank and Gumbel copula but the abilities are the same for the two halves.
Model C is the most flexible one, as it allows for differences between the two halves resulting to a much larger number of parameters.  Table \ref{prel} displays the results. Model B, i.e.\ the model with a Frank copula and the same parameters in the two halves is the preferred one. The Frank copula may be preferred over the Gumbel as we do not observe any tail dependence in the data. 

\begin{table}
\begin{tabular}{llccc}
Model & &log-likelihood & \#par & AIC \\
\hline
Model A  && -1590.047  & 72 & 3324.09\\
Model B  \\
&Gumbel & -1603.235 & 38 & 3282.47 \\
&Frank  & -1601.736 & 38& {\bf 3279.47} \\
Model C \\
&Gumbel & -1587.511  & 73 & 3321.02\\
&Frank & -1586.922& 73 & 3319.84  \\
\hline
\end{tabular}
\caption{\label{prel} Log-likelihood of fitted models of different complexity.}
\end{table}

\begin{center}
\begin{figure}
\centering
\includegraphics[scale=0.5]{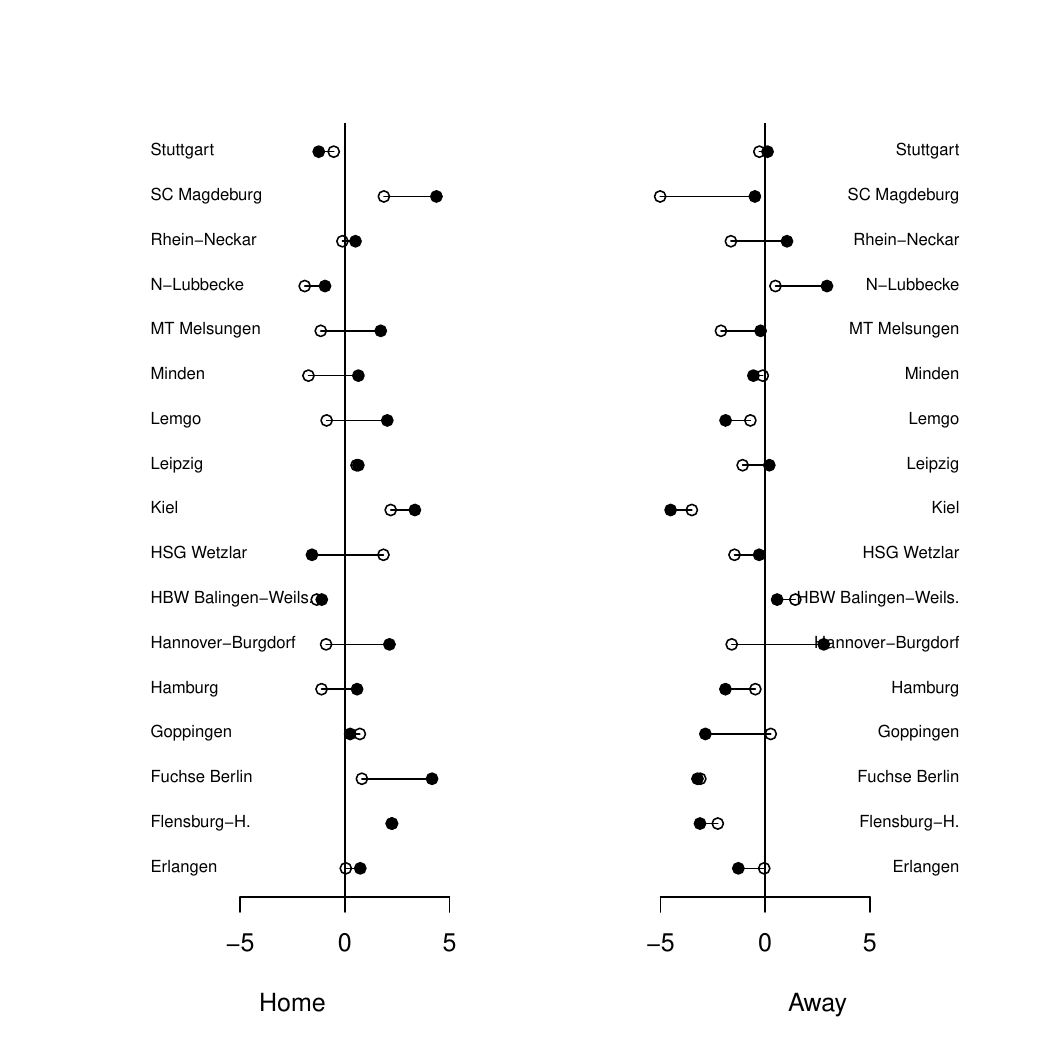}
\caption{\label{homeaway}  Estimated parameters for the 1$^{\text{st}}$  and 2$^{\text{nd}}$  half for home and away parameters. For most of the teams, the estimated parameters are quite close. First half estimates are indicated by the solid points, 2$^{\text{nd}}$ half estimates are given by the open points.}
\end{figure}
\end{center}

The results support the parsimonious model that the parameters representing the abilities of the teams in home and away matches are the same. To further investigate this finding, Figure \ref{homeaway} shows the estimated parameters for the first and the second half. One can see that for most of the teams the parameters for the first and second half are very close supporting that we can be more parsimonious by considering common parameters.

\subsection{Conditional distributions}

In this section, we consider the case that we know the half time result and we want to find the probabilities for the final score difference. 
Based on the fitted bivariate model, we can calculate the conditional probabilities
\[
P(Y=y | X=x) = \frac{P(Y=y,X=x)}{P(X=x)},
\]
where the nominator is a bivariate Skellam through a copula and the denominator is the marginal Skellam distribution. 
From this we calculate the probability for winning the game at the end. So if  team, whose goals are denoted by $Y$, is trailing by say $x$ goals at half time, then  the probability of winning is 
\[
P(\text{Win}) = P(Y > -x | X=x) = \sum\limits_{y= -x+1}^\infty P(Y > x | X=x)
\]

\begin{center}
\begin{figure}
\centering
\includegraphics[scale=0.4]{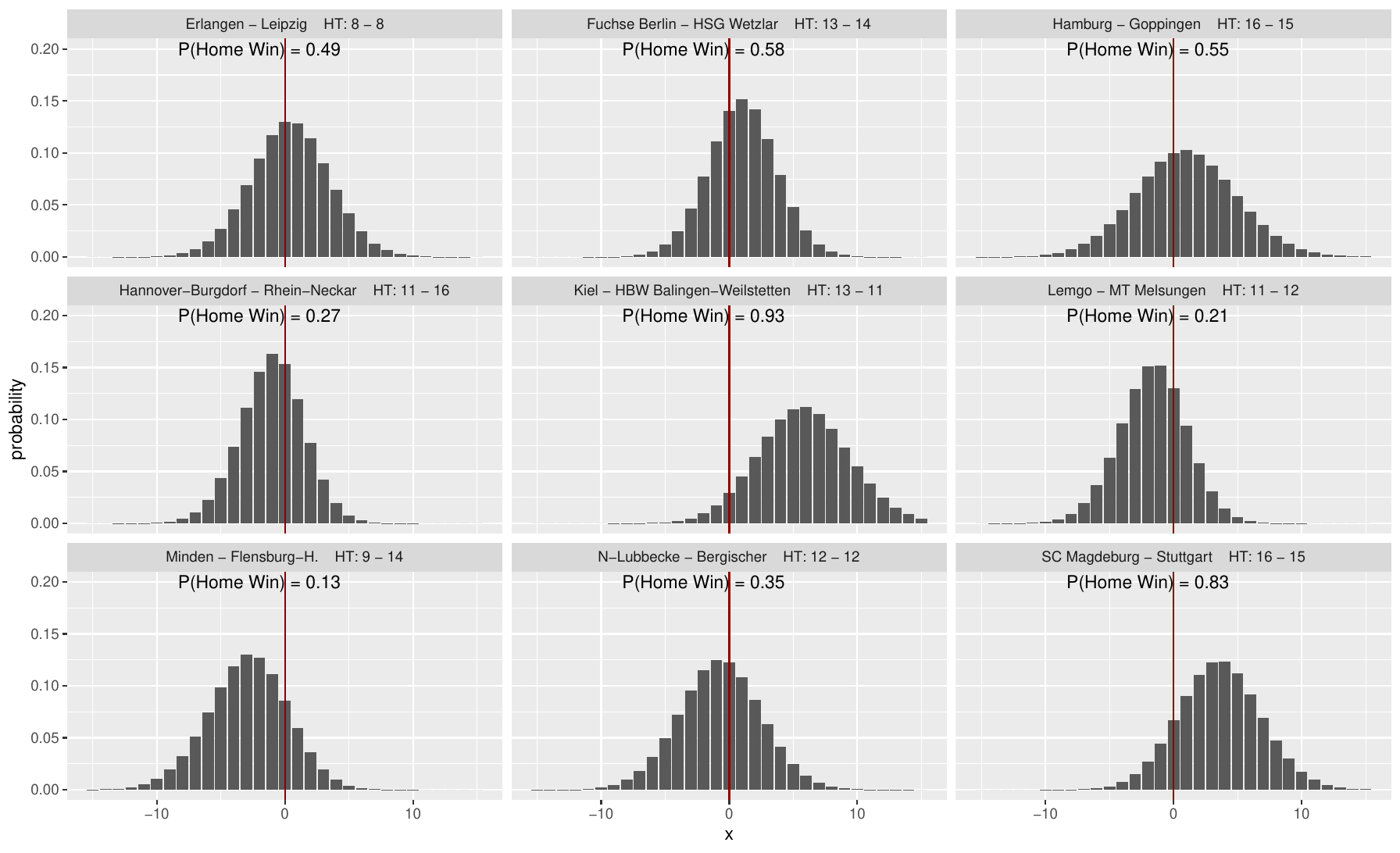}
\caption{\label{final}  Conditional distribution for $Y|X=x$, for 
all the matches of the first day. The plots show the conditional distribution given the half time score for the difference in the final score. We also report the probability that the home team  to win.
The half time score can be read in the caption of each plot. The vertical line refers to the probability of a draw.}
\end{figure}
\end{center}

 Figure \ref{final} presents the conditional probabilities that the home team will win at the end of the game conditional on the score of the first half, for the first matchday of the 2021-2022 season. For the prediction of these probabilities, we have used the model that was selected in Table \ref{prel}.
 So, for example, looking at the top left plot, we see that at the half time the score between Erlangen and Leipzig was simply 8-8, and based on the team abilities estimated we calculated that the probability that Erlangen will win is 0.49, while the entire conditional distribution can be also seen. 
 
\section{Discussion}

In this study, the outcomes of handball results were modelled in several ways. Initially, we decomposed the bivariate pair of a result by transforming it into a univariate statistic, i.e.\ the score difference. We used the Skellam distribution as the baseline model for modelling this difference. Considering the patterns found in the data --- an excess of draws --- we considered a zero-inflated version of the baseline model. Additionally, we compared both models with a discretised normal and Laplace distributions. 
Modelling the score difference overcomes the problem to specify the distributions for the number of goals, which is not easy since we need underdispersed discrete distributions. 

In a second step, we addressed the problem of modelling the score differences of the two halves in a handball match separately. Here, we employed copulas to model the dependence between both halves' score differences. Such bivariate model formulations can be used to predict the final outcome as a conditional probability, i.e.\ based on the score difference at half time. 

The models provided in the present paper can serve as the basis for potential further model extensions. For example, they can be the basis for modelling certain characteristics of the game, such as the ``7 vs 6'' rule introduced in 2016, where a team can substitute the goalkeeper to have more players in offence. This is a risky situation that has not been examined in detail in the literature. To investigate the effect of replacing the goalkeeper for an additional outfield player, one could consider the models developed here --- in particular, to estimate the effect on the score difference associated with such strategy.

Finally, note that betting, and in particular in-game betting, is also an important component of sports \cite{michels2023bettors, otting2024demand}), also in handball. 
Since in handball the score changes too often we need a model that can adjust on the score differential and the remaining time in a flexible way. The Skellam distribution shares such a property in the sense that changing the time domain the distribution is still in the same family with altered parameters and hence it can be very helpful for predicting in-game winning probabilities.

\section{Competing Interests}
The authors have no competing interests to declare that are relevant to the content of this article.

\bibliography{sn-bibliography}

\begin{thebibliography}{}
\providecommand{\doi}[1]{\url{https://doi.org/#1}}
\bibcommenthead

\bibitem[\protect\citeauthoryear{Abramowitz and Stegun}{Abramowitz and
  Stegun}{1974}]{mABR70a}
Abramowitz, M. and I.A. Stegun. 1974.
\newblock {\em Handbook of {M}athematical {F}unctions}.
\newblock New York: Dover.

\bibitem[\protect\citeauthoryear{Barbiero}{Barbiero}{2014}]{barbiero2014alternative}
Barbiero, A. 2014.
\newblock An alternative discrete skew {L}aplace distribution.
\newblock {\em Statistical Methodology\/}~16: 47--67 .

\bibitem[\protect\citeauthoryear{Bryson, Dolton, Reade, Schreyer, and
  Singleton}{Bryson et~al.}{2021}]{bryson2021causal}
Bryson, A., P.~Dolton, J.J. Reade, D.~Schreyer, and C.~Singleton. 2021.
\newblock Causal effects of an absent crowd on performances and refereeing
  decisions during {C}ovid-19.
\newblock {\em Economics Letters\/}~198: 109664 .

\bibitem[\protect\citeauthoryear{Csat{\'o}}{Csat{\'o}}{2021}]{csato2021coronavirus}
Csat{\'o}, L. 2021.
\newblock Coronavirus and sports leagues: obtaining a fair ranking when the
  season cannot resume.
\newblock {\em IMA Journal of Management Mathematics\/}~{\em 32\/}(4): 547--560
  .

\bibitem[\protect\citeauthoryear{Dixon and Coles}{Dixon and
  Coles}{1997}]{dixon1997modelling}
Dixon, M.J. and S.G. Coles. 1997.
\newblock Modelling association football scores and inefficiencies in the
  football betting market.
\newblock {\em Journal of the Royal Statistical Society: Series C (Applied
  Statistics)\/}~{\em 46\/}(2): 265--280 .

\bibitem[\protect\citeauthoryear{Dumangane, Rosati, and Volossovitch}{Dumangane
  et~al.}{2009}]{dumangane2009departure}
Dumangane, M., N.~Rosati, and A.~Volossovitch. 2009.
\newblock Departure from independence and stationarity in a handball match.
\newblock {\em Journal of Applied Statistics\/}~{\em 36\/}(7): 723--741 .

\bibitem[\protect\citeauthoryear{Felice}{Felice}{2023}]{felice2023ranking}
Felice, F. 2023.
\newblock Ranking handball teams from statistical strength estimation.
\newblock {\em arXiv preprint arXiv:2307.06754\/} .

\bibitem[\protect\citeauthoryear{Felice and Ley}{Felice and
  Ley}{2023}]{felice2023prediction}
Felice, F. and C.~Ley. 2023.
\newblock Prediction of handball matches with statistically enhanced learning
  via estimated team strengths.
\newblock {\em arXiv preprint arXiv:2307.11777\/} .

\bibitem[\protect\citeauthoryear{Groll, Heiner, Schauberger, and
  Uhrmeister}{Groll et~al.}{2020}]{groll2020prediction}
Groll, A., J.~Heiner, G.~Schauberger, and J.~Uhrmeister. 2020.
\newblock Prediction of the 2019 {IHF} {W}orld {M}en’s {H}andball
  {C}hampionship--a sparse {G}aussian approximation model.
\newblock {\em Journal of Sports Analytics\/}~{\em 6\/}(3): 187--197 .

\bibitem[\protect\citeauthoryear{Holmes and McHale}{Holmes and
  McHale}{2024}]{holmes2024forecasting}
Holmes, B. and I.G. McHale. 2024.
\newblock Forecasting football match results using a player rating based model.
\newblock {\em International Journal of Forecasting\/}~{\em 40\/}(1): 302--312
  .

\bibitem[\protect\citeauthoryear{Irwin}{Irwin}{1937}]{irwin1937frequency}
Irwin, J.O. 1937.
\newblock The frequency distribution of the difference between two independent
  variates following the same {P}oisson distribution.
\newblock {\em Journal of the Royal Statistical Society\/}~{\em 100\/}(3):
  415--416 .

\bibitem[\protect\citeauthoryear{Jiang, Mao, and Wu}{Jiang
  et~al.}{2014}]{jiang2014skellam}
Jiang, L., K.~Mao, and R.~Wu. 2014.
\newblock A {S}kellam model to identify differential patterns of gene
  expression induced by environmental signals.
\newblock {\em BMC Genomics\/}~{\em 15\/}(1): 1--9 .

\bibitem[\protect\citeauthoryear{Karlis and Mamode~Khan}{Karlis and
  Mamode~Khan}{2023}]{karlis2023models}
Karlis, D. and N.~Mamode~Khan. 2023.
\newblock Models for integer data.
\newblock {\em Annual Review of Statistics and its Application\/}~10: 297--323
  .

\bibitem[\protect\citeauthoryear{Karlis and Ntzoufras}{Karlis and
  Ntzoufras}{2006}]{KarlisNtzoufras06SIM}
Karlis, D. and I.~Ntzoufras. 2006.
\newblock Bayesian analysis of the differences of count data.
\newblock {\em Statistics in Medicine\/}~{\em 25\/}(11): 1885--1905 .

\bibitem[\protect\citeauthoryear{Koopman, Lit, and Lucas}{Koopman
  et~al.}{2017}]{koopman2017intraday}
Koopman, S.J., R.~Lit, and A.~Lucas. 2017.
\newblock Intraday stochastic volatility in discrete price changes: the dynamic
  {S}kellam model.
\newblock {\em Journal of the American Statistical Association\/}~{\em
  112\/}(520): 1490--1503 .

\bibitem[\protect\citeauthoryear{Lago-Pe{\~n}as, Gomez, Via{\~n}o, and
  Gonz{\'a}lez-Garc{\'\i}a}{Lago-Pe{\~n}as et~al.}{2013}]{lago2013home}
Lago-Pe{\~n}as, C., M.A. Gomez, J.~Via{\~n}o, and I.~Gonz{\'a}lez-Garc{\'\i}a.
  2013.
\newblock Home advantage in elite handball: the impact of the quality of
  opposition on team performance.
\newblock {\em International Journal of Performance Analysis in Sport\/}~{\em
  13\/}(3): 724--733 .

\bibitem[\protect\citeauthoryear{Maher}{Maher}{1982}]{maher1982modelling}
Maher, M.J. 1982.
\newblock Modelling association football scores.
\newblock {\em Statistica Neerlandica\/}~{\em 36\/}(3): 109--118 .

\bibitem[\protect\citeauthoryear{McHale and Scarf}{McHale and
  Scarf}{2007}]{mchale2007modelling}
McHale, I. and P.~Scarf. 2007.
\newblock Modelling soccer matches using bivariate discrete distributions with
  general dependence structure.
\newblock {\em Statistica Neerlandica\/}~{\em 61\/}(4): 432--445 .

\bibitem[\protect\citeauthoryear{Michels, {\"O}tting, and Karlis}{Michels
  et~al.}{2023}]{michels2023extending}
Michels, R., M.~{\"O}tting, and D.~Karlis. 2023.
\newblock Extending the {D}ixon and {C}oles model: an application to women's
  football data.
\newblock {\em arXiv preprint arXiv:2307.02139\/} .

\bibitem[\protect\citeauthoryear{Michels, {\"O}tting, and Langrock}{Michels
  et~al.}{2023}]{michels2023bettors}
Michels, R., M.~{\"O}tting, and R.~Langrock. 2023.
\newblock Bettors’ reaction to match dynamics: Evidence from in-game betting.
\newblock {\em European Journal of Operational Research\/}~{\em 310\/}(3):
  1118--1127 .

\bibitem[\protect\citeauthoryear{Ntzoufras, Palaskas, and Drikos}{Ntzoufras
  et~al.}{2021}]{ntzoufras2021bayesian}
Ntzoufras, I., V.~Palaskas, and S.~Drikos. 2021.
\newblock Bayesian models for prediction of the set-difference in volleyball.
\newblock {\em IMA Journal of Management Mathematics\/}~{\em 32\/}(4): 491--518
  .

\bibitem[\protect\citeauthoryear{{\"O}tting, Michels, Langrock, and
  Deutscher}{{\"O}tting et~al.}{2024}]{otting2024demand}
{\"O}tting, M., R.~Michels, R.~Langrock, and C.~Deutscher. 2024.
\newblock Demand for live betting: An analysis using state-space models.
\newblock {\em Applied Stochastic Models in Business and Industry\/} .

\bibitem[\protect\citeauthoryear{Pelechrinis and Winston}{Pelechrinis and
  Winston}{2021}]{pelechrinis2021skellam}
Pelechrinis, K. and W.~Winston. 2021.
\newblock A {S}kellam regression model for quantifying positional value in
  soccer.
\newblock {\em Journal of Quantitative Analysis in Sports\/}~{\em 17\/}(3):
  187--201 .

\bibitem[\protect\citeauthoryear{Prieto, G{\'o}mez, Volossovitch, and
  Sampaio}{Prieto et~al.}{2016}]{prieto2016effects}
Prieto, J., M.{\'A}. G{\'o}mez, A.~Volossovitch, and J.~Sampaio. 2016.
\newblock Effects of team timeouts on the teams’ scoring performance in elite
  handball close games.
\newblock {\em Kinesiology\/}~{\em 48\/}(1.): 115--123 .

\bibitem[\protect\citeauthoryear{Singh, Scarf, and Baker}{Singh
  et~al.}{2023}]{singh2023unified}
Singh, A., P.~Scarf, and R.~Baker. 2023.
\newblock A unified theory for bivariate scores in possessive ball-sports: The
  case of handball.
\newblock {\em European Journal of Operational Research\/}~{\em 304\/}(3):
  1099--1112 .

\bibitem[\protect\citeauthoryear{Skellam}{Skellam}{1946}]{jg1946frequency}
Skellam, J. 1946.
\newblock The frequency distribution of the difference between two {P}oisson
  variates belonging to different populations.
\newblock {\em Journal of the Royal Statistical Society. Series A\/}~{\em
  109\/}(3): 296--296 .

\bibitem[\protect\citeauthoryear{Smiatek and Heuer}{Smiatek and
  Heuer}{2012}]{smiatek2012statistical}
Smiatek, J. and A.~Heuer. 2012.
\newblock A statistical view on team handball results: home advantage, team
  fitness and prediction of match outcomes.
\newblock {\em arXiv preprint arXiv:1207.0700\/} .

\bibitem[\protect\citeauthoryear{Tomy and Veena}{Tomy and
  Veena}{2022}]{tomy2022retrospective}
Tomy, L. and G.~Veena. 2022.
\newblock A retrospective study on {S}kellam and related distributions.
\newblock {\em Austrian Journal of Statistics\/}~{\em 51\/}(1): 102--111 .

\bibitem[\protect\citeauthoryear{Van~Eetvelde, Hvattum, and Ley}{Van~Eetvelde
  et~al.}{2023}]{van2023probabilistic}
Van~Eetvelde, H., L.M. Hvattum, and C.~Ley. 2023.
\newblock The probabilistic final standing calculator: a fair stochastic tool
  to handle abruptly stopped football seasons.
\newblock {\em AStA Advances in Statistical Analysis\/}~{\em 107\/}(1-2):
  251--269 .

\bibitem[\protect\citeauthoryear{Volossovitch and Debanne}{Volossovitch and
  Debanne}{2021}]{volossovitch2021home}
Volossovitch, A. and T.~Debanne. 2021.
\newblock Home advantage in handball.
\newblock {\em Home Advantage in Sport: Causes and the Effect on
  Performance\/}: 220--227 .

\end{thebibliography}

\end{document}